# Bose-Einstein condensation of photons in an optical microcavity

Jan Klaers, Julian Schmitt, Frank Vewinger, and Martin Weitz

*Institut für Angewandte Physik, Universität Bonn, Wegelerstr. 8, 53115 Bonn, Germany*

**Bose-Einstein condensation, the macroscopic ground state accumulation of particles with integer spin (bosons) at low temperature and high density, has been observed in several physical systems, including cold atomic gases and solid state physics quasiparticles[1-9]. However, the most omnipresent Bose gas, blackbody radiation (radiation in thermal equilibrium with the cavity walls) does not show this phase transition. Here the photon number is not conserved when the temperature of the photon gas is varied (vanishing chemical potential),[10] and at low temperatures photons disappear in the cavity walls instead of occupying the cavity ground state. Theoretical works have considered photon number conserving thermalization processes, a prerequisite for Bose-Einstein condensation, using Compton scattering with a gas of thermal electrons[11], or using photon-photon scattering in a nonlinear resonator configuration[12,13]. In a recent experiment, we have observed number conserving thermalization of a two-dimensional photon gas in a dye-filled optical microcavity, acting as a 'white-wall' box for photons[14].**

**Here we report on the observation of a Bose-Einstein condensation of photons in a dye-filled optical microcavity. The cavity mirrors provide both a confining potential and a non-vanishing effective photon mass, making the system formally equivalent to a two-dimensional gas of trapped, massive bosons. By multiple**

**scattering of the dye molecules, the photons thermalize to the temperature of the dye solution (room temperature). Upon increasing the photon density we observe the following signatures for a BEC of photons: Bose-Einstein distributed photon energies with a massively populated ground state mode on top of a broad thermal wing, the phase transition occurring both at the expected value and exhibiting the predicted cavity geometry dependence, and the ground state mode emerging even for a spatially displaced pump spot. Prospects of the observed effects can include studies of extremely weakly interacting low-dimensional Bose gases[9] and new coherent ultraviolet sources[15].**

Fifty years ago, the invention of the laser has provided us with a source of coherent light. In a laser, optical gain is achieved under conditions where both the state of the gain medium and the state of the light field are far removed from thermal equilibrium[16]. The realization of a light source with a macroscopically populated photon mode not being the consequence of a laser-like gain, but rather owing to an equilibrium phase transition of photons has so far been prevented by the lack of a suitable number conserving thermalization process[17]. For strongly coupled mixed states of photons and excitons (exciton polaritons), a thermalization process induced by interparticle interactions between excitons has been reported to lead to a (quasi-)equilibrium Bose-Einstein condensation of polaritons[5-7]. In the present work, rapid decoherence due to frequent collisions of dye molecules with the solvent prevents coherent energy exchange between photons and dye molecules and therefore the condition of strong matter-field coupling is not met[18,19]. We therefore can assume the relevant particles to be well described by photons instead of polaritons.

Our experiment confines photons in a curved-mirror optical microresonator filled with a dye solution, where photons are repeatedly absorbed and reemitted by the dye molecules. The small distance of 3.5 optical wavelength between the mirrors causes a large frequency spacing between adjacent longitudinal modes (free spectral range $\cong 7 \times 10^{13} Hz$), comparable with the spectral width of the dye, see Fig.1a, and modifies spontaneous emission such that the emission of photons with a given longitudinal mode number, q=7 in our case, dominates over other emission processes. By this the longitudinal mode number is frozen out and the remaining transverse modal degrees of freedom make the photon gas effectively two-dimensional. Moreover, the dispersion relation becomes quadratic (i.e. non-relativistic), as indicated in Fig.1b, with the frequency of the q=7 transverse ground mode (TEM$_{00}$) acting as a low-frequency cutoff with $\omega_{cutoff} \cong 2\pi \cdot 5.1 \times 10^{14} Hz$. The curvature of the mirrors induces a harmonic trapping potential for the photons (see Methods). This is indicated in Fig.1c, along with a scheme of the experimental setup.

Thermal equilibrium of the photon gas is achieved by absorption and reemission processes in the dye solution, which acts as heat bath and equilibrates the transverse modal degrees of freedom of the photon gas to the (rovibrational) temperature of the dye molecules (see Methods). The photon frequencies will accumulate within a range $k_B T / \hbar (\cong 2\pi \cdot 6.3 \times 10^{12} Hz$ at room temperature) above the low-frequency cutoff. In contrast to the case of a blackbody radiator, for which the photon number is determined by temperature (Stefan-Boltzmann law), the number of (optical) photons in our microresonator is not altered by the temperature of the dye solution, as purely thermal excitation is suppressed by a factor of order $\exp(\hbar\omega_{cutoff}/k_B T) \approx \exp(-80)$. The thermalisation process thus conserves the average photon number.

Our system is formally equivalent to an ideal gas of massive bosons with an effective mass $m_{ph} = \hbar\omega_{cutoff}/c^2 \cong 6.7 \times 10^{-36} kg$ moving in the transverse resonator plane, harmonically confined with a trapping frequency $\Omega = c/\sqrt{D_0 R/2} \cong 2\pi \cdot 4.1 \times 10^{10} Hz$ (see Methods), with c as the speed of light in the medium, $D_0 \cong 1.46 \mu m$ the mirror separation and $R \cong 1m$ the radius of curvature. A harmonically trapped two-dimensional ideal gas exhibits BEC at finite temperature[20,21], in contrast to the two-dimensional homogeneous case. We correspondingly expect a Bose-Einstein condensation when the photon wave packets spatially overlap at low temperatures or high densities, i.e. the phase space density $n\lambda_{th}^2$ exceeds a value near unity. Here $n$ denotes the number density, photons per area, and $\lambda_{th} = h/\sqrt{2\pi m_{ph} k_B T} \cong 1.58 \mu m$ (defined in analogy to e.g. a gas of atoms[17]) the de Broglie wavelength associated with the thermal motion in the resonator plane. Note that $\lambda_{th} = 2\sqrt{\pi}/k_{r,rms}$, with $k_{r,rms} = \sqrt{\langle k_r^2 \rangle_T}$ as the rms transverse component of the photon wavevector at temperature $T$. The precise onset of Bose-Einstein condensation in the two-dimensional, harmonically trapped system can be determined from a statistical description using a Bose-Einstein distributed occupation of trap levels[14,20,21], giving a critical particle number of

$$N_c = \frac{\pi^2}{3}\left(\frac{k_B T}{\hbar\Omega}\right)^2. \tag{1}$$

At room temperature (T = 300 K), we arrive at $N_c \cong 77000$. It is interesting to note that both the thermal energy $k_B T$ and the trap level spacing $\hbar\Omega$ are roughly a factor $10^9$ above the corresponding values in atomic physics BEC experiments[2-4], however the ratio $k_B T/\hbar\Omega \cong 150$, corresponding to the mean excitation value per axis, is quite comparable.

By pumping the dye with an external laser we add to a reservoir of electronic excitations that exchanges particles with the photon gas, in the sense of a grand-canonical ensemble. The pumping is maintained throughout the measurement to compensate for losses due to coupling into unconfined optical modes, finite quantum efficiency and mirror losses. In a steady state, the average photon number will be $N_{ph} = N_{exc} \cdot \tau_{ph} / \tau_{exc}$, where $N_{exc}$ is the number of molecular excitations, $\tau_{exc}$ is their electronic lifetime in the resonator (of order of a nanosecond) and $\tau_{ph} \cong 20 ps$ is the average time between emission and reabsorption of a photon. For a detailed description of the thermalisation, it is important to realize that it originates from particle exchange with a reservoir being in equilibrium. The latter is characterized by rovibrational molecular states that are highly equilibrated both in the lower and in the upper electronic level due to subpicosecond relaxation[22] from frequent collisions with solvent molecules. This process efficiently decorrelates the states of absorbed and emitted photons, and leads to a temperature dependent absorption and emission spectral profile that is responsible for the thermalization. To relax both spatially and spectrally to an equilibrium distribution, a photon has to scatter several times of molecules before being lost. In previous work, we have shown that the photon gas in the dye-filled microcavity system can be well described by a thermal equilibrium distribution, showing that photon loss is sufficiently slow[14]. To avoid excessive population of dye molecule triplet states and heat deposition, the pump beam is acoustooptically chopped to $0.5\mu s$ pulses, which is at least two orders of magnitude above the described timescales, with $8ms$ repetition time.

Typical room temperature spectra for increasing pumping power are given in Fig.2a (recorded using rhodamine 6G dye solved in methanol, $1.5 \times 10^{-3} M$). At low pumping and correspondingly low intracavity power we observe a spectrally broad emission, which is in good agreement with a room temperature Boltzmann distribution of photon energies above the cavity cutoff[14]. With increasing pump power the maximum of the spectral distribution shifts towards the cavity cutoff, i.e. it more resembles a Bose-Einstein distribution function. For a pumping power above threshold, a spectrally sharp peak at the frequency of the cavity cutoff is observed, while the thermal wing shows saturation. The described signatures are in good agreement with theoretical spectra based on Bose-Einstein distributed transversal excitations (inset of Fig.2a). At the phase transition the power inside the resonator is $P_{c,\exp} = (1.55 \pm 0.60)W$, corresponding to a critical photon number of $(6.3 \pm 2.4) \times 10^4$. This value still holds when rhodamine is replaced by perylene-diimide (PDI) solved in acetone ($0.75 g/l$), i.e. for both dyes the measured critical number is in agreement with the value predicted for a Bose-Einstein condensation of photons (eq.1).

Spatial images of the photon gas below and slightly above criticality are shown in Fig.2b. In either case the lower energetic (yellow) photons are bound to the trap centre while the higher energetic (green) photons appear at the outer trap regions. Above the critical photon number a bright spot is visible in the trap centre with a (FWHM) diameter of $(14 \pm 3) \mu m$, indicating a macroscopically populated TEM$_{00}$-mode (expected diameter $12.2 \mu m$). Fig.2c shows normalized spatial intensity profiles along one axis for increasing pumping power near the critical value. Interestingly, we observe that the mode diameter enlarges with increasing condensate fraction, as shown in Fig.2d. This effect is not expected for an ideal gas of photons. In principle, this could be due to a Kerr-nonlinearity in the dye solution, but the most straightforward explanation is a weak repulsive optical self-interaction from thermal lensing in the dye (which can be modelled by a mean-field interaction, see methods). From the increase of the mode diameter we can estimate the magnitude of this effective repulsive interaction, yielding a dimensionless interaction parameter of $\tilde{g} \approx (7 \pm 3) \times 10^{-4}$. This is much below the values $\tilde{g} = 10^{-2} \ldots 10^{-1}$ reported for two-dimensional atomic physics quantum gas experiments and also below the value at which Kosterlitz-Thouless (KT) physics, involving "quasi-long-range" order, is expected to become relevant[23]. The latter is supported by an experiment directing the condensate peak through a shearing interferometer, where we have not seen signatures of the phase blurring expected for a KT-phase[24].

We have tested for a dependence of the BEC threshold on the resonator geometry. From eq.1 we expect a critical optical power $P_c = (\pi^2/12)(k_B T)^2 (\omega_{cutoff}/\hbar c) R$, which grows linearly with the mirror radius of curvature R and is independent of the longitudinal mode number q. Fig.3a and the lower panel of Fig.3b show corresponding measurements of the critical power, with results in good agreement with both the expected absolute values and the expected scaling. The upper panel of Fig.3b gives the required optical pump power to achieve the phase transition versus the number of longitudinal modes, showing a decrease because of stronger pump power absorbance for larger mirror spacing. This is in strong contrast to results reported from "thresholdless" dye-based optical microlasers, for which an *increase* of the threshold pumping power was observed[25,26]. For a macroscopic laser a fixed value of the pump intensity is required to reach inversion.

Finally, we have investigated the condensation for a spatially mismatched pumping spot. Due to the thermal redistribution of photons we expect that even a spatially displaced pump beam can provide a sufficiently high photon density at the trap centre to reach the phase transition. This effect is not known in lasers, but is observed in the framework of polariton condensation[7]. For our measurement, the pump beam (diameter $\approx 35 \mu m$) was displaced at $\approx 50 \mu m$ distance from the trap centre. Fig.4 shows a series of spatial intensity profiles recorded for a fixed pumping power and for different values of the cavity cut-off wavelength $\lambda_{cutoff}$, which tunes the degree of thermalization[14]. The lower graph gives results recorded with $\lambda_{cutoff} \cong 610 nm$, for which the maximum of the fluorescence is centred at the position of the pump spot (shown by a dashed line). The weak reabsorption in this wavelength range prevents efficient photon thermalization. When the cutoff is moved to shorter wavelengths, the stronger reabsorption in this wavelength range leads to increasingly symmetric photon distributions around the trap centre. For a cavity cut-off near $570 nm$ we observe a small bright spot at the position of the TEM$_{00}$-mode. The corresponding cusp in the intensity profile of Fig. 4 indicates a condensate fraction of roughly $N_0 / N \approx 1\%$. These measurements show that due to the photon thermalization, Bose-Einstein condensation can be achieved even when the pumping intensity at the position of the ground state mode is essentially zero.

To conclude, evidence for a Bose-Einstein condensation of photons was obtained from (i) the spectral distribution that shows Bose-Einstein distributed photon energies including a macroscopically occupied ground state, (ii) the observed threshold of the phase transition, which shows both the predicted absolute value and scaling with resonator geometry, (iii) condensation, which appears at the trap centre even for a spatially disjunct pump spot. It is instructive to discuss the relation of a photon BEC to microlasers, which also use high finesse cavities to capture the emission of excited state atoms and molecules in a small volume[25-27]. The low lasing thresholds and potentially inversion-less oscillation of microlasers however result from a high coupling efficiency of spontaneous photons into a single cavity mode – which is not the case in the work reported here, as is evident from the observed highly multimodal emission below criticality. The main borderline between a laser and Bose-Einstein condensation remains that the latter device (in contrast to the former) operates in thermal equilibrium, with the macroscopically populated mode being a consequence of equilibrium Bose statistics.

An interesting consequence of a grand-canonical particle exchange between photon gas and the reservoir of excited state dye molecules, is that unusually large number fluctuations of the condensed phase could occur[28]. We expect that the concept of photon condensation holds promise for the exploration of novel states of light, and for light sources in new wavelength regimes.

**Methods**

**Preparation of photon gas.**   Photons are confined in an optical microresonator consisting of two curved dielectric mirrors with >99.997% reflectivity, filled with dye solution. The dye is pumped with a laser beam near $532 nm$ wavelength ($\approx 70 \mu m$ diameter except for Fig.4) directed under $45^0$ to the optical axis, exploiting a reflectivity minimum of the mirrors. The trapped photon gas thermalizes to the rovibrational temperature T of the dye solution by repeated absorption and emission processes, as follows from a detailed balance condition[14] fulfilled in media obeying the Kennard-Stepanov relation $f_T(\omega)/\alpha_T(\omega) \propto \omega^3 e^{-\hbar\omega/k_B T}$, which describes a temperature dependent frequency scaling of absorption coefficient $\alpha_T(\omega)$ versus emission strength $f_T(\omega)$[29]. Noteworthy, this equilibrium between photons and the dye solution is reminiscent of Einstein's description of the heat contact between radiation and a Doppler-broadened gas[30].

**Photon dispersion in cavity and optical self-interaction.**   The photon energy in the resonator as a function of transversal ($k_r$) and longitudinal ($k_z$) wavenumber reads $E = \hbar c\sqrt{k_z^2 + k_r^2}$, where c denotes the speed of light in the medium. The boundary conditions yield $k_z(r) = q\pi/D(r)$, where $D(r) = D_0 - 2(R - \sqrt{R^2 - r^2})$ gives the mirror separation at distance $r$ from optical axis. For fixed longitudimal mode number q and in paraxial approximation ($r \ll R$, $k_r \ll k_z$), one arrives at the dispersion of a particle with nonvanishing mass $m_{ph} = \hbar k_z(0)/c = \hbar\omega_{cutoff}/c^2$, with its motion restricted to the (two-dimensional) transverse resonator plane under harmonic confinement with trapping frequency $\Omega = c\sqrt{2/D_0 R}$, see[14]. A possible self-interaction of photons (Kerr lensing or thermal lensing in the limit of neglibile transverse heat flow) can be incorporated by $n(r) = n_0 + n_2 I(r)$, where $I(r)$ is the optical intensity and $n_0 (\cong 1.33$ for methanol) and $n_2$ are the linear and nonlinear indices of refraction respectively, yielding

$$E \cong m_{ph}c^2 + \frac{(\hbar k_r)^2}{2m_{ph}} + \frac{1}{2}m_{ph}\Omega^2 r^2 - m_{ph}c^2 \frac{n_2}{n_0}I(r). \tag{2}$$

The latter term resembles a mean-field interaction familiar from the Gross-Pitaevskii equation for atomic BECs, which, using a dimensionless interaction parameter[23] $\tilde{g} := -(m_{ph}^4 c^6 n_2)/(2\pi\hbar^3 n_0 q)$ and wavefunction $\psi(r)$ with $I(r) = (m_{ph}c^2)^2 (hq)^{-1} N_0 |\psi(r)|^2$, can be written in the more familiar form

$$E_{int} = (\hbar^2/m_{ph})\tilde{g}N_0|\psi(r)|^2.$$

**Acknowledgements**

We thank J. Dalibard and Y. Castin for helpful discussions. Financial support from the Deutsche Forschungsgemeinschaft within the focused research unit FOR557 is acknowledged. M. W. thanks the IFRAF for support of a guest stay at LKB Paris, where part of the given discussion on interacting two-dimensional photon gases has been developped.

**Figure captions**

**Figure 1**: Cavity mode spectrum and setup. **a**, Schematic spectrum of cavity modes and (relative) absorption coefficient and fluorescence strength of Rhodamine 6G dye versus frequency. Transverse modes belonging to the manifold of longitudinal mode number q=7 are shown by black lines, those of other longitudinal mode numbers in grey. The degeneracy of a given transversal energy is indicated by the height of the bars. **b**, Dispersion relation of photons in the cavity (solid line), with fixed longitudinal mode number (q=7), and the free photon dispersion (dashed line). **c**, Scheme of the experimental setup. The trapping potential for photons imposed by the curved mirrors is indicated on the left hand side.

**Figure 2**: Spectral and spatial intensity distribution. **a**, Spectral intensity distributions (connected circles) transmitted through one cavity mirror, as measured with a spectrometer, for different pump powers. The intracavity power (in units of $P_{c,\exp} = (1.55 \pm 0.60)W$) is derived from the power transmitted through one cavity mirror. A spectrally sharp condensate peak at the cavity cutoff is observed above a critical power level, with a width limited by the spectrometer resolution. The inset gives theoretical spectra (solid lines) based on a Bose-Einstein distribution of photons for different particle numbers at room-temperature[14]. **b**, Images of the spatial radiation distribution transmitted through one cavity mirror both below (upper panel) and above (lower panel) criticality, showing a macroscopically occupied $TEM_{00}$-mode for the latter case. **c**, Cut through the centre of the intensity distribution for increasing optical pump powers and, **d**, width of the condensate peak versus condensate fraction along with a theoretical model based on the Gross-Pitaevskii equation with an interaction parameter $\tilde{g} = 7 \times 10^{-4}$ (Methods). Error bars are the systematic uncertainties. (*q=11* for figures c,d. All other measurements use *q=7*).

**Figure 3**: Critical power. **a**, Intracavity power at criticality for different curvatures of the cavity mirrors. The dashed line shows the theoretical expectation based on eq. 1. **b**, Intracavity power at criticality (lower panel) versus longitudinal mode number q. The upper panel shows the required optical pump power $P_{pump,c}$ (circles) along with a fit $P_{pump,c} \propto (q-q_0)^{-1}$ yielding $q_0 = 4.77(25)$. For this we assume an inverse proportionality to the absorption length in the dye $q-q_0$, where $q_0$ incorporates an effective penetration depth into the cavity mirrors. The above value for $q_0$ is in good agreement with an independent measurement of the pump power transmission, yielding $q_0 = 4.68(17)$. Error bars are systematic uncertainties.

**Figure 4**: Spatial redistribution of photons. Intensity profiles recorded with a pump beam spot (diameter $\approx 35 \mu m$) spatially displaced by $50 \mu m$ from the trap centre, for different cutoff wavelengths. For a cutoff wavelength of $610 nm$ (bottom profile), where reabsorption is weak, the emitted radiation is centered at the position of the pump spot, whose profile is shown by the dashed line (measured by removing one of the cavity mirrors). When tuning the cavity cutoff to shorter wavelength values, where the reabsorption efficiency is increased, light is redistributed towards the trap centre. For data recorded with $570 nm$ cutoff wavelength (top profile), a cusp appears, corresponding to a partly condensed state with a condensate fraction $N_0/N$ of about 1%.

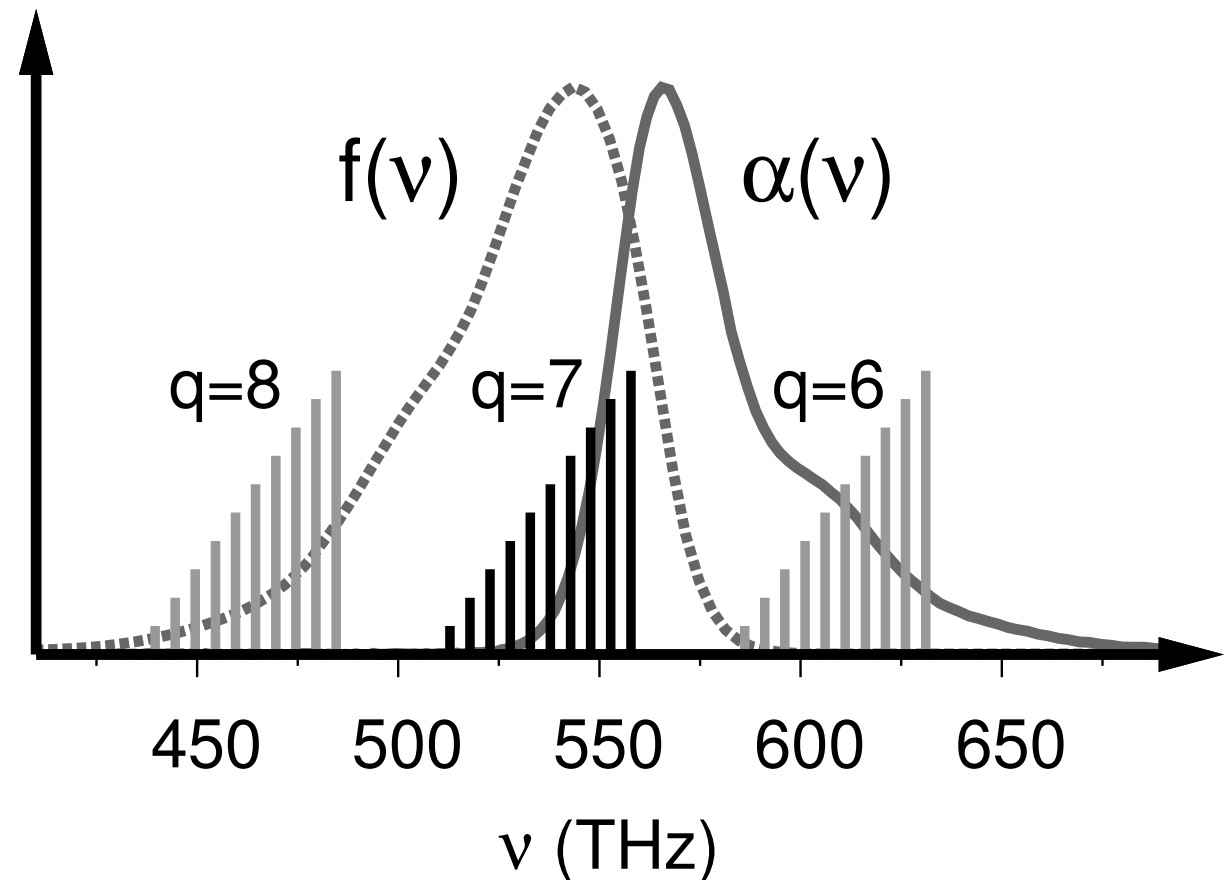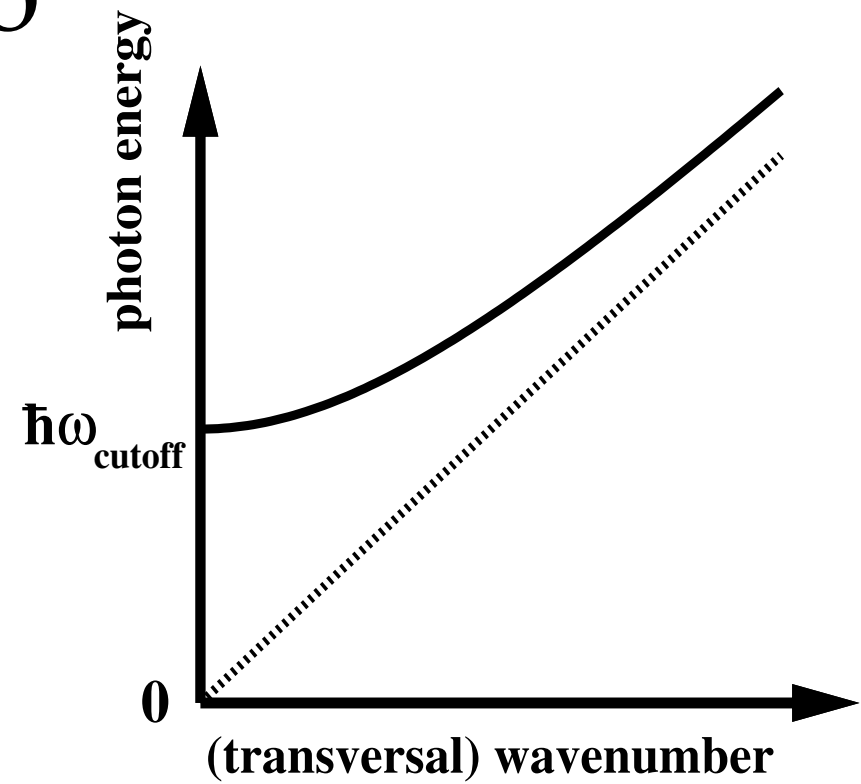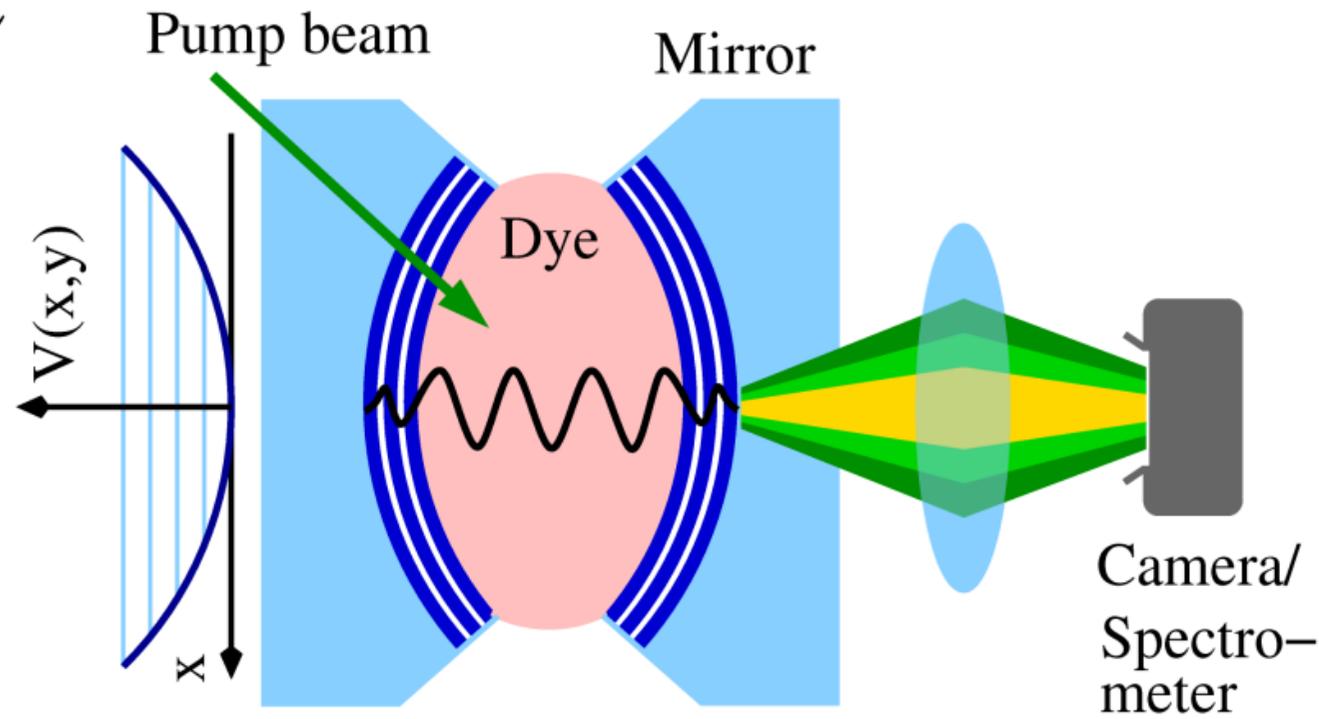

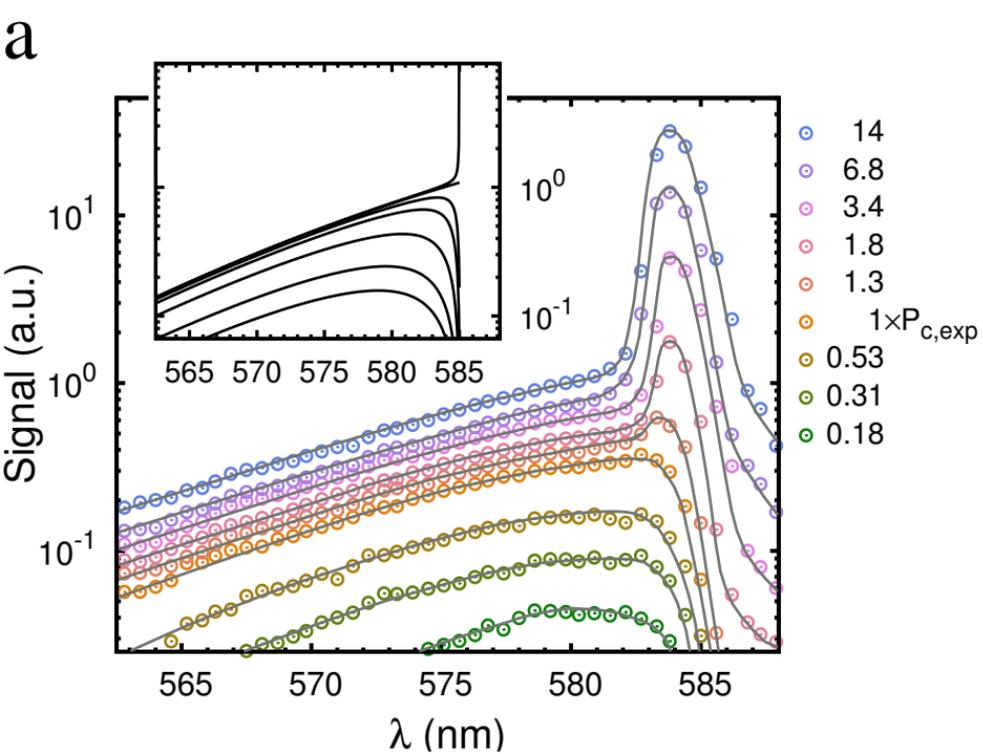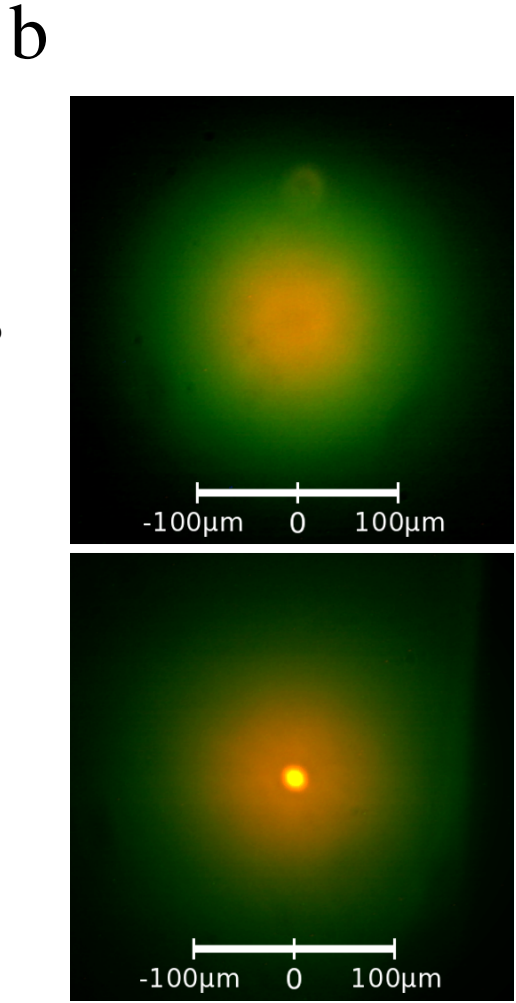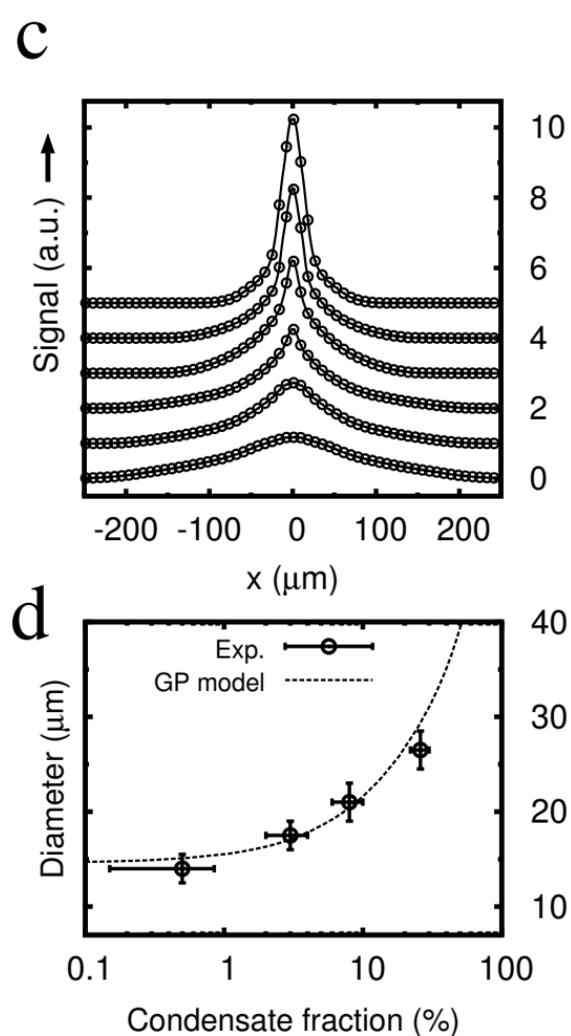

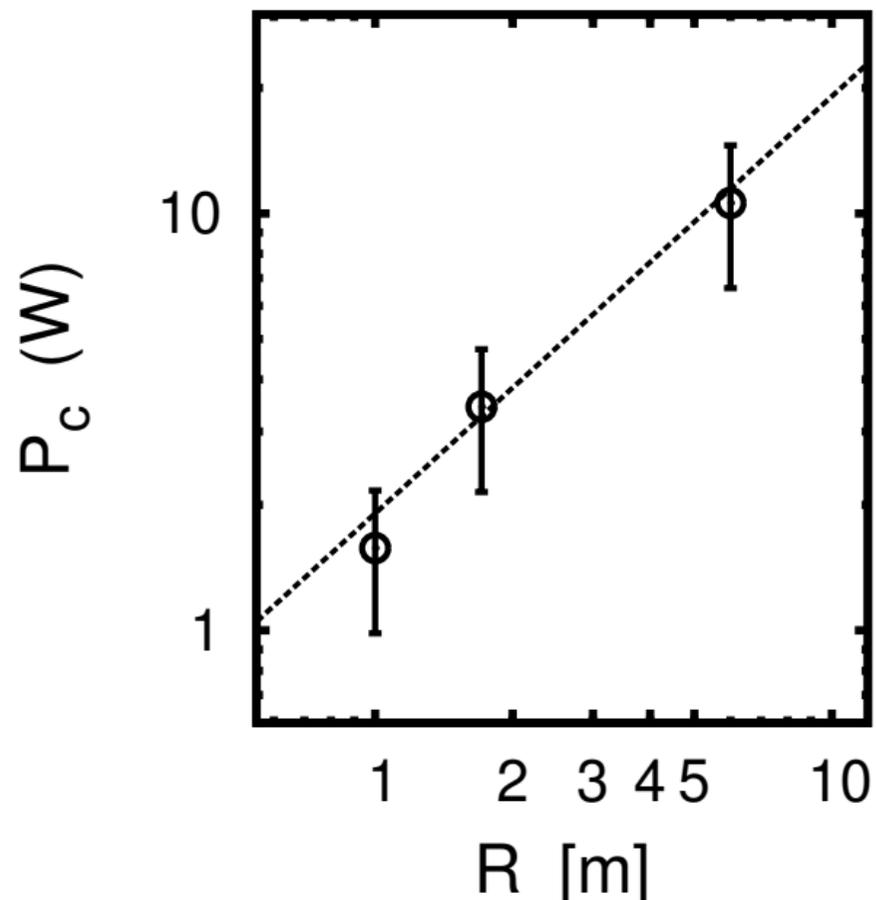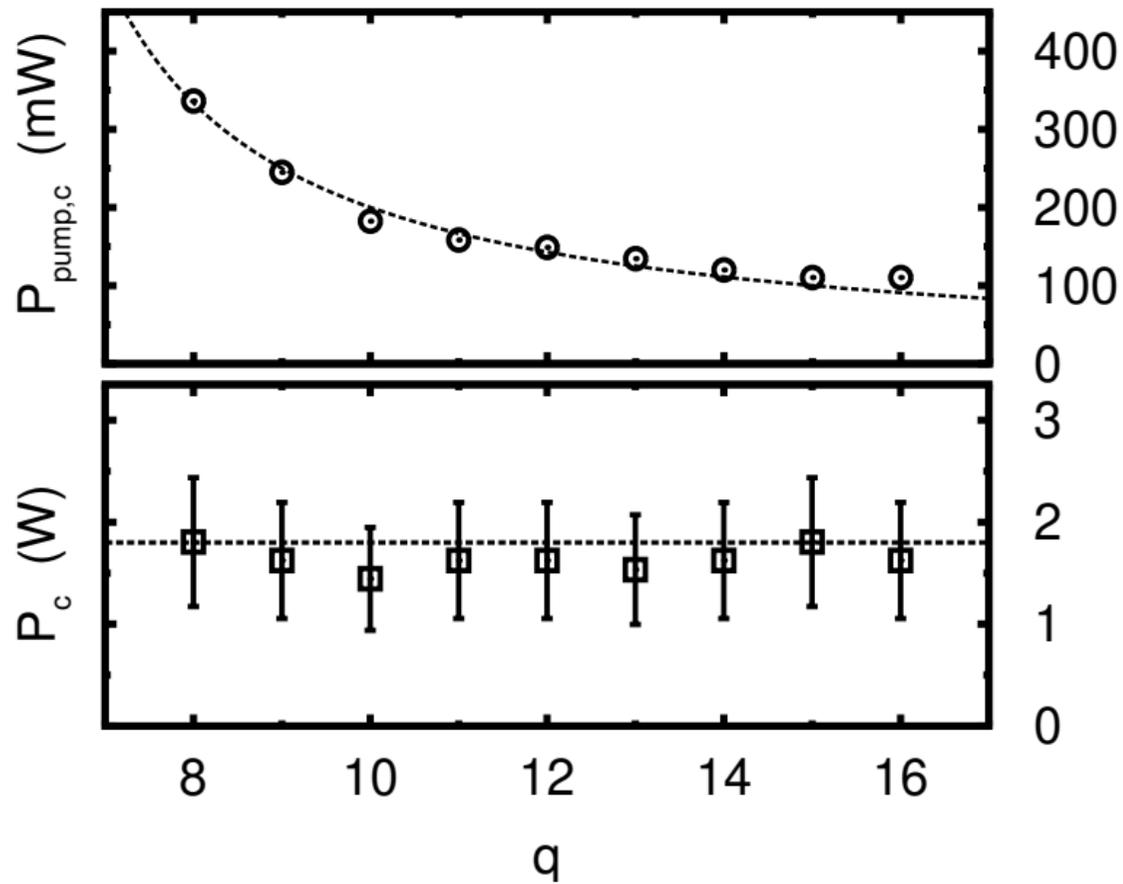

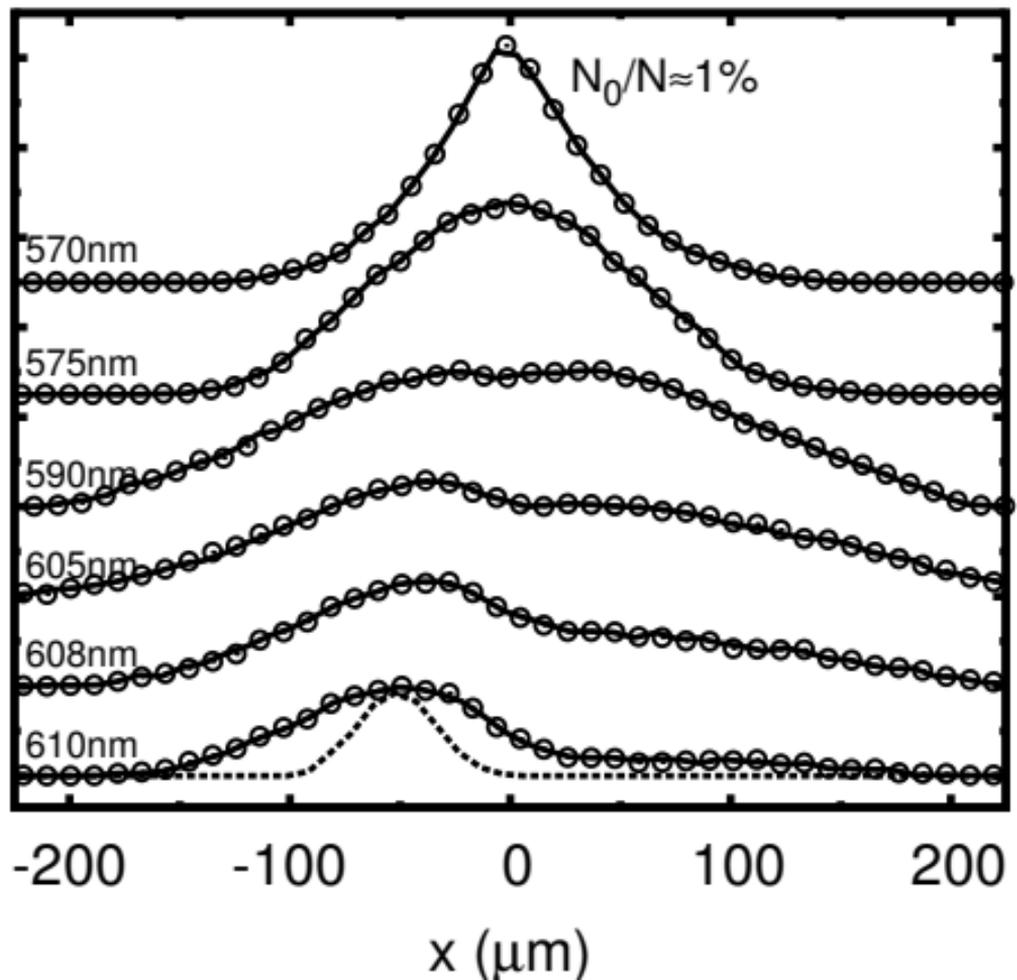